# Refounding legitimacy towards Aethogenesis

Olivier Auber

Evolution, Complexity and COgnition (ECCO lab), Vrije Universiteit Brussel (VUB).
Global Brain Institute. P2P foundation.

Olivier.auber@vub.ac.be

**Abstract:** The fusion of humans and technology takes us into an unknown world, described by some authors as populated by quasi-living species that would relegate us—ordinary humans—to the rank of alienated agents, emptied of our identity and consciousness. I argue instead that our world is woven of simple, though invisible, perspectives, which—if we become aware of them—may renew our ability for making judgments and enhance our autonomy. I became aware of these invisible perspectives by observing and practicing a real-time collective net art experiment called the Poietic Generator. As the perspectives unveiled by this experiment are invisible, I have called them anoptical perspectives (i.e. non-optical), by analogy with the optical perspective of the Renaissance. Later, I have come to realize that these perspectives obtain their cognitive structure from the political origins of our language. Accordingly, it is possible to define certain cognitive criteria for assessing the legitimacy of the anoptical perspectives, just like some artists and architects of the Renaissance defined the geometrical criteria that established the legitimacy of the optical one.

___

I will not say a word about the first singularity of the universe (the alleged Big Bang). I will not say much about the second singularity that took place on earth and probably on many other planets: the explosion of biological codes called abiogenesis, which realized 'the transition of a world without biology to a world with biology'. Most species that emerged during the abiogenesis disappeared or mutated, while a small number have proliferated into the present.

However, let me express an interest in the third singularity: The origination of human language, and especially what could lead to the fourth one, as a consequence of the possible inception of two classes of artificial living species: Artificial General Intelligence (AGI) and Mindplexes (Yudkowsky, Goertzel, 2003).

## The political origin of language

The evolutionary anthropologist Cadell Last has recently described a condition he calls atechnogenesis (Last 2015) as 'the passage of a world without technology to a world with technology', which we are experiencing at the moment on earth and which that may have also occurred on other planets. According to him, the atechnogenesis will lead to the birth of technological life on a cosmological scale.

Following Claude Lévi-Strauss and André Leroi-Gourhan, we can easily imagine that one can trace the atechnogenesis on our planet back to the emergence of the genus Homo itself; that is to say, to the emergence of a proto-language among some hominins in conjunction with the inception of the first tools.

The cognitive scientist Jean-Louis Dessalles specifies the link between language and technologies. He puts forward strong arguments that suggest that the invention of weapons was the singularity that triggered the explosion of our symbolic codes (Dessalles 2014a, 2014b).

According to his theory, weaponry in its most prehistoric version would have made the social order based on physical domination brutally obsolete. For the first time, weapons would have allowed hominins to kill without risk, not only wild beasts, but, above all: Dominant peers.

*Once, for whatever reason, easy killing became possible among our hominin ancestors, the absolute right of the strongest instantaneously became obsolete.* Jean-Louis Desssalles.

This political crisis induced by the armed threat of the dominated against the dominants would have put enormous stress on our species, to the point of becoming a factor of natural selection. The individuals whose behavior could have contributed to the collective survival would have been selected. The selected individuals would have been those who proved their capability to both identify unexpected signs of danger to the community and to communicate them to their peers by hand gestures, vocalizations, and later by an increasingly articulated language. As analyzed by Dessalles, language would have been our Evolutionarily Stable Strategy (ESS), which allowed us to escape the political crisis triggered by the invention of weapons. The appearance of human language would have overthrown the brute force to become our main daily activity, as well as the driving force that shapes our social structure.

**The fourth singularity?**

If weapons were the factor that triggered the emergence of human language, both continued to evolve together until they were almost inseparable. This is so much so the case now that we live within their complex entanglement.

Even more refined than military weapons, I believe that the mother of all weapons lies within the language itself. It is invisible, immaterial, and essentially logical. That is money.

While the primordial weapons were used by dominated peers in order to threaten the dominant ones, contrariwise, the monetary weapon is exclusively in the hands of the dominants (who should not be confused with the elites, as we shall see). Money is not only a social code that promotes one class over another; by driving the technological development, it becomes an instrument of total domination. Indeed, capital–traditionally seen as the accumulation of stocks and means of production—has been suddenly dematerialized in the form of complex economic and symbolic values activated by robots and artificial intelligence working for an ever-smaller number of people, i.e. the 'vectorialist class', which owns the vectors of our interactions (Wark 2004). In its hands, the monetary weapon is both the goal and the tool of a global cyber warfare.

Who cannot see that this exponential concentration leads to the creation of vast desert areas, where currency, as well as other economic and symbolic values, are in the best cases maintained under perfusion? The monetary code, based on its mode of creation by debt (Graeber 2013), appears more and more like a carnivorous operating system that inevitably leads to destruction. It is as if its unrestricted expansion allows a kind of quasi-living system to practice a systemic predation of our species. That is to say, it can safely, blindly, and indiscriminately kill our fellow humans (as well as animals), whatever their social ability, to the point that the ancient order, based on our language, seems to have become obsolete.

Our species seems in the process of being submitted to a new political stress, similar to the one that led to the invention of language, but in the opposite direction:

What is this phenomenon of such a magnitude if it is not the announcement of a new singularity? Many researchers think that the economic crisis summarizes all other crises and that it drives us to a singularity. Some of them are considering a radical rebuilding of the monetary system on the foundation of a monetary creation process based on a universal dividend (Laborde, 2010). Some others imagine how to match offer and demand without using any currency (Heylighen 2015a). There are uncountable other proposals of alternative currency models (ACM), all of which are possible seeds for new monetary orders. Some might be beneficial, but others could also reboot an even worse domination.

**Artificial Currency Models as artificial living species**

If these ACMs are potential new weapons, we should control them a priori from a certain perspective. But in practice, Distributed Autonomous Organizations (DAO) emulated by ACMs are popping up everywhere like spontaneous generations of artificial living species within a transitioning ecology. As they try to proliferate on the same playing field of our economic relations, they compete with one another. We know that they will combine, mutate, or disappear until a new, revitalized ecosystem emerges.

Among the different scenarios that I will explore, is perhaps hided the pathway of our future Evolutionarily Stable Strategy. The question arises, how to detect it, build it, and select it? I will argue that this is all a question of legitimacy.

**Anoptical perspectives as a cognitive model**

To assist us, we can try to observe a fact that has gone relatively unnoticed since the beginning of the Industrial Revolution, and particularly since the emergence of telecommunications. It is the inception of two invisible perspectives that are driving the topology of our networks. Since these perspectives are invisible, I propose to call them anoptical perspectives (i.e. non-optical) by analogy with the optical perspective of the Renaissance (Auber 2001).

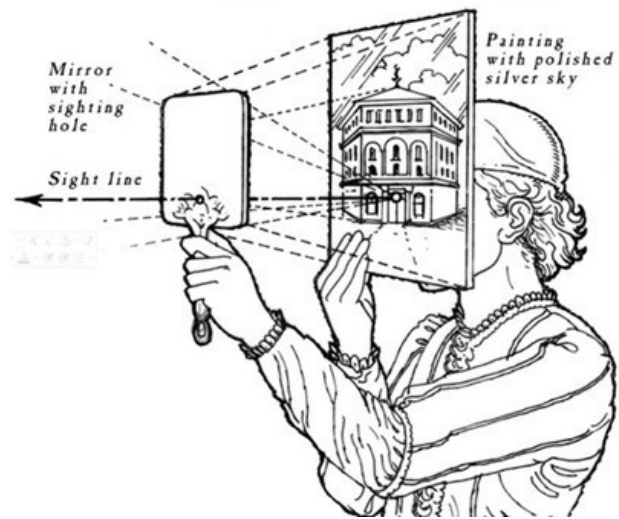

Filippo Brunelleschi (1377–April 15, 1446).
Two panel paintings illustrating for the first time around 1413 the geometry of the optical perspective.

Just as Brunelleschi's experiment demonstrated the concept of optical perspective at the beginning of the Renaissance, the observation and the practice of a very simple net art experiment called the Poietic Generator[1] may unfold the anoptical ones.

The Poietic Generator allows a large group of people to perform a real-time collective interaction over a network. The experiment can run either on a centralized or on a distributed peer-to-peer network. In both cases, the Poietic Generator achieves seemingly the same kind of human interaction. A feedback loop between the individuals and the group produces an emergence of unpredictable shapes that can be seen and interpreted by all, and in which everyone can take action.

The Poietic Generator may be seen as a generic model of multiple complex systems such as informational, financial, urban, or ecological networks in which everyone is involved daily. But unlike these 'actual' networks, the rules and infrastructure of which are often opaque, the Poietic Generator is perfectly transparent. Everything is known or knowable, in particular the fact that it operates according to a centralized or a distributed structure. Indeed, the networks that determine every economic and symbolic value can operate according to these two architectures. The reasons why the centralized architecture is chosen in the majority of the cases will be discussed below.

---
1 Poietic Generator : http://poietic-generator.net

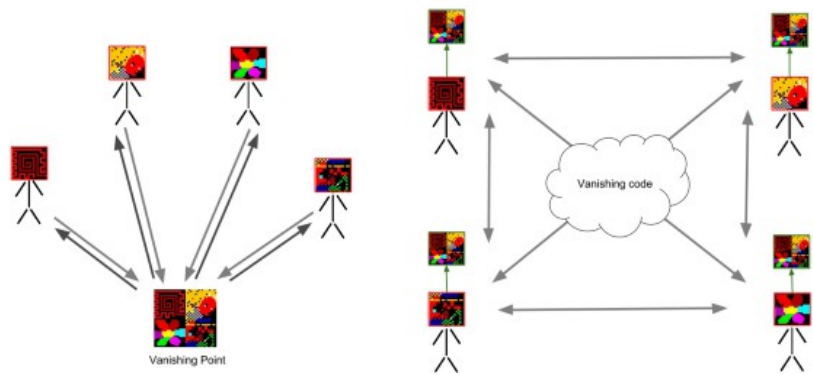

Poietic generator's centralized or distributed network.

In the case of a centralized network architecture, one can speak about a temporal perspective (TP). Indeed, the center is the physical place where the network's subjective time emerges moment by moment from the interaction among its members. One can call this physical place a 'temporal vanishing point'. For example, that could be a server.

In the case of a distributed peer-to-peer architecture, one can speak about a digital perspective (DP). An arbitrary digital code guarantees the emergence of the network's own subjective time in each of its nodes. One can call this code a 'vanishing code'. That could be, for example, a multicast IP address, a hashtag or a blockchain, and in the case of a biological network, it could be a molecule, a genome, a DNA sequence, etc.

The two anoptical perspectives (TP, DP), unveiled by the Poietic Generator over two different types of networks, share numerous topological and symbolic attributes with the optical perspective.

In both network architectures, the vanishing point/code is the symbol of the infinite and the unknowable. Indeed, even if it is perfectly defined and/or located (the server is somewhere, the source code of the network is accessible and readable), the emergence of the global image/phenomena remains unpredictable.

The vanishing point/code is also homologous to a certain point of view, which is analogous to the painter's eye in the optical perspective. This point of view defines the set of rules of the network. In the case of the Poietic Generator's experiment, this is simply me.

At last, as the optical perspective is working according to some geometrical rules, the anoptical perspectives are working according to cognitive ones. In this regard, it is possible to define some criteria of legitimacy for the building of anoptical perspectives, just as painters and architects defined the legitimate optical perspective during the Renaissance.

The legitimacy of the anoptical perspectives can be observed in the conditions that determine the feedback loop between the local agents and the global emergence. One can consider as a first approximation that—if the feedback loop is thoroughly achieved without any external manipulation; that is if each agent of the network can observe its own input among the other ones—then it is possible to pretend that the anoptical perspective is legitimate. In that case, one can trace this legitimacy back from the patterns that emerge from the vanishing point/code. The global network behaves like an autopoietic system with an operational closure (Maturana &and Varela 1992); that is, a pure living organism.

On the contrary, the legitimacy of the anoptical perspective is not achieved in the case where some external manipulations somehow break or disturb the feedback loop by injecting some alien data or modifying, hiding, or retaining some parts of the agent's inputs. In that case, we can say that the agents are 'alienated' in the sense defined by Heinz Von Foerster, which is, in short:

*They cannot recognize or observe their own trace among the whole.*

One can trace the manipulation back in the patterns designed by the system, even though it is difficult because the network behaves 'like' a living organism; that is an organism that mimics life according to a certain model of life injected by the author(s) of the manipulation (consciously or not).

Feedback manipulations are called by my friend Florence Meichel: Noloops. She finds Noloops in every corner of human activity (industry, media, finance, politics, religion, science, and, above all, 'social networks'). This is not a surprise. Indeed, in a context where 'third order structural couplings' (Maturana & and Varela 1992) between systems are everywhere, Noloops are the norm. Practicing social networks (which are in fact interconnected 'social silos') show this very well. Some Noloops are well known and somehow tolerable, while some others are hidden and ethically highly problematic. Think about Volkswagen's cheating algorithms.

Upon this basis, let me try to define more precisely the notion of legitimacy of the anoptical perspectives, not in philosophical, moral, political, or legal terms, but in cognitive ones.

**Three criteria of legitimacy (A, AB, ABC)**

Noloops refer to questions of cheating and autonomy. These notions can be considered in the framework of Dessalles' model of human language, which I take the risk of summarizing succinctly in a single sentence:
Signalling unexpected events allows the individuals to strengthen their social network in order to reduce their chances of being killed by surprise.
Some peers choose an 'honest signalling' strategy. Their main quality is to detect inconsistencies or to restore consistency. The peers who show this ability, among other qualities, are accepted as members of the group and, step by step, they constitute its 'elite'.

On the other hand, some other individuals choose to cheat in order to get a dominant position in a less costly manner. Basic cheaters are the ones who over-signal irrelevant events to get some attention or advantage from their peers. The latter are easily neutralized by the ones who can 'restore consistency', just by damaging their reputation.

There exist of course highly various and uncountable cheat strategies that are far more difficult to detect. Countermeasures are also numerous. Nevertheless, one thing has not been considered in Dessalles' framework so far. Cheating strategies may apply to the vectors of the signalling activity ; nowadays, that is networks and technology in general. Yet, we can try to classify these cheating strategies in the framework of the anoptical perspectives.

Cheating can intervene at three main levels of these perspectives: (1) peer level, (2) network level, and (3) vanishing point/code level. Here are some examples taken from contemporary practices:

> 1) Peer level
> Over-signalling, fake, noise injection, high frequency trading, multiple identities, identity theft, obfuscation, virus, groupthink, harassing, etc.
>
> 2) Network level
> Spying, real-time filtering, deep packet inspection, backdoor, intentional network outage, manipulation of counters and social graphs, etc.
>
> 3) Vanishing point/code level
> Institutional bias, nudge, fiction, Ponzi scheme, lying algorithms, irrelevant model injection, lobbying, star system, monopoly, takeover, putsch, paradigmatic blindness, etc.

All these cheating strategies applied to the vectors of our signalling activity tend to deprive us of our feedback or to mislead us. Since we are 'alienated', even if we have a great ability to 'restore consistency', we cannot be recognized as friends by the others (and eventually given access to a position of eliteness). In the ecological niches defined by these cheating activities, dominant positions are definitely taken by cheaters (who are cheating consciously or not).

The globalization of these 'vectorial cheating strategies' gives us a cognitive explanation for the new political stress that we are facing as a species in this critical stage of atechnogenesis. These strategies let the dominants not only practice a systematic predation over the dominated, but let them definitively neutralize the elites, which could/should restore the consistency. That leads to an overall failure of consistency, a global mismatch between the behavior of our organization and the state of its environment.

In these dramatic conditions, peers have to urgently find some countermeasures for deactivating the 'vectorial cheating' in order to recover their autonomy and their capability of judgment.
Here we propose three basic criteria that could help agents/peers/users to evaluate the legitimacy of any given centered or distributed network (TP or DP) to which it/he/she belongs:

> A) Does any agent A have the actual right to access the network if he requests it? Can A leave the network freely?
>
> AB) Is any agent B (present or future, including agents that conceive, administer, and develop the network) treated like A?
>
> ABC) If agents A, B, and C (where ABC is the beginning of a multitude) belong to a network that meets the first two criteria, are they peers? That is, are they able to recognize, trust, and respect each other, thereby building common representations and common sense?

In other words, the A criterion aims to avoid any form of hostage taking. People are free to come and go. The networks that they practice are appreciated under this criterion for their 'exchange value' and their 'use value'. Contrariwise, the networks that are irreversible for technical reasons, or those that are built to maximize their 'hostage value' for speculative and predatory purposes, are removed.
The AB criterion offers to evaluate the symmetry of the network using an approach of reciprocity. Networks whose design is mechanically producing concentrations of power are abandoned.

ABC is a more complex criterion. It proposes to consider the notion of 'peer' without giving any definitive answer. For example, is it enough that a peer takes part in my network for looking at him as a peer? On the contrary, can I nevertheless consider someone who does not belong to my network as my peer? Does a peer have to be necessarily human? Could it be non-human? If so, what should be the ratio human/non-human? Etc. In the era of atechnogenesis, nobody will escape the need to answer these questions, and it is likely that answers will be contingent and scalable.

Nevertheless, the beginning of a generic response is given by the anoptical perspectives. Indeed, this ABC criterion suggests ultimately to question the imaginary projections that we make on networks and other technological artifacts. We know that, in a certain way, they play the roles of 'transitional objects' (Winnicott), 'instinctual objects' (Freud) or 'object a' (Lacan). In terms of anoptical perspectives, we can say that we cannot avoid to project onto their vanishing point/code our desire, our need of a symbol that connects us to the Other and differentiates us from it. As has already been said, these anoptical vanishing points/codes are completely unknowable. Consequently, the shared mourning of their mastery and their knowledge constitutes paradoxically the common base of our social links.

In reality, we try by all means to grasp them, just like in paintings of the Baroque era, where the vanishing point of the optical perspective is obscured by the Christ, the Virgin Mary, or the dove of the Holy Spirit. Today we superimpose some abstract notions and some political slogans on the vanishing points/codes of the networks, according to the case: 'market', 'performance', 'competitiveness', 'communication', etc. Maybe tomorrow we will use concepts such as 'extension of life' or 'eternal life'. In fact, these slogans exclude the peers who are not within 'the market', who are neither 'effective' nor 'competitive', those who know how to live without the illusion of 'communication' or 'eternal life'. Therefore, the network selects the peers. But as we will see further, the strict respect of the anoptical legitimacy should be the opposite. The peers should select the networks.

To conclude this part. The above-mentioned criteria of legitimacy (A, AB, ABC) are in line with many other attempts of mankind to restore some consistency. They propose a reformulation appropriated to networks of some ancient and somewhat forgotten principles, for example, those of the French Republic: LIBERTY—EQUALITY—FRATERNITY.

They are also expanding the principles of free software (Stallman R. 1983) to a whole information system.

**Four evolutionary scenarios**

As is well known, about two centuries were required for the literacy of the optical perspective to spread around the world after its inception by Brunelleschi in the early 1400s, which triggered enormous effects. It quickly became the main tool that we use to project ourselves in time and space; that is, to evaluate distance, the time/energy needed to travel across it, as well as the resources that may be gathered. This literacy was certainly the relevant intellectual tool for an era of conquerors, explorers, and builders.

This is no longer the case, as has been announced by the deconstruction of the concept of space made by major artists of the twentieth century. Anoptical perspectives that are driving the development of networks definitely prevail upon the optical one which organizes our spatial perception.

Unfortunately, despite the warnings, the optical perspective still remains the current imaginary mode of our society. In particular, this representation mode, which is inconsistent with the very nature of our environment, directly affects the current monetary system. Indeed, everything happens as if our social relationships as well as our minds were considered as territories that can be conquered, occupied, and exploited. In this regard, the notion of emotional capitalism (Pierre, Alloing 2015)—which is the total industrialization of emotions—seems to constitute a relevant description of this dramatic drift.

The anoptical perspective's literacy could offer a countermeasure to that drift. To understand the phenomena it may trigger, let us consider the research of Dessalles again. By following his basic assumptions about the origin and the function of human language, Dessalles deduces a mathematical model of our signaling activity that has been validated by actual data. In short, the level of our signaling activity and our personal attractiveness are linked in a surprising manner. One can observe three classes of people on Dessalles' diagrams:

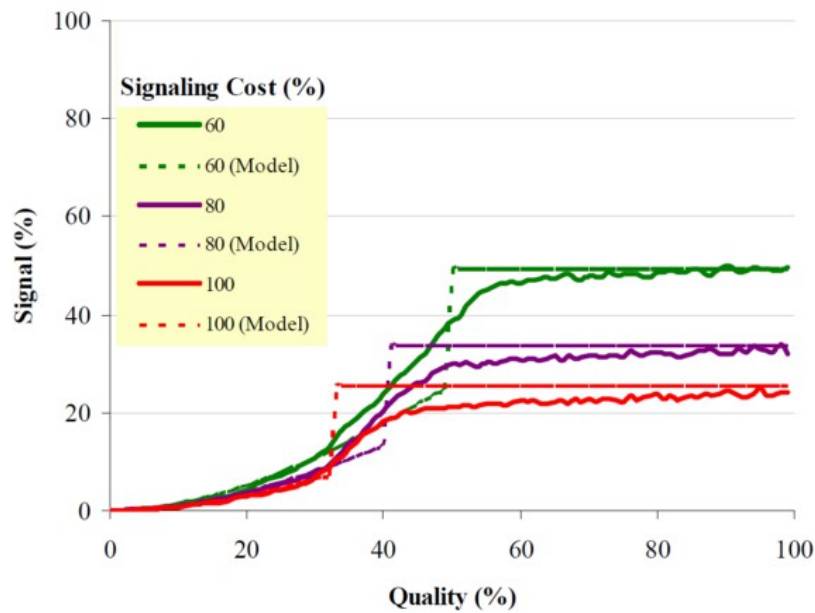

The Y-axis measures the level of the personal signaling.
The X-axis measures the 'quality' of individuals, that is, the attractiveness that determines the size of their social network. (in Dessalles J-L 2014b, 'Optimal Investment in Social Signals', pp. 9, Evolution, 68(6), 1640–1650).

The left side of the diagram includes individuals who have a low attractiveness and get only a limited social network. These individuals are showing a low signaling activity.

The central part of the diagram includes individuals who are competing with each other through their signaling activity in order to increase their personal attractiveness and thus the size of their own social network.

In the right side of the diagram, the high personal attractiveness of individuals has almost no impact on their signalling activity, which remains limited in comparison to their social status. Individuals in this group are very slightly competing with each other.

Consequently, this modeling which is directly deduced from the characteristics of our language, shows how far the shape of our social hierarchy is itself an aspect of our ESS.

From that moment on, the question is whether the global industrialization (vectorization) of our overall social signalling behavior implemented by centralized or decentralized networks has any impact on our ESS, and whether the awareness of the anoptical perspectives may produce a slight change?
On the brink of the fourth singularity, let me propose below four evolutionary scenarios of our current ESS for discussion:

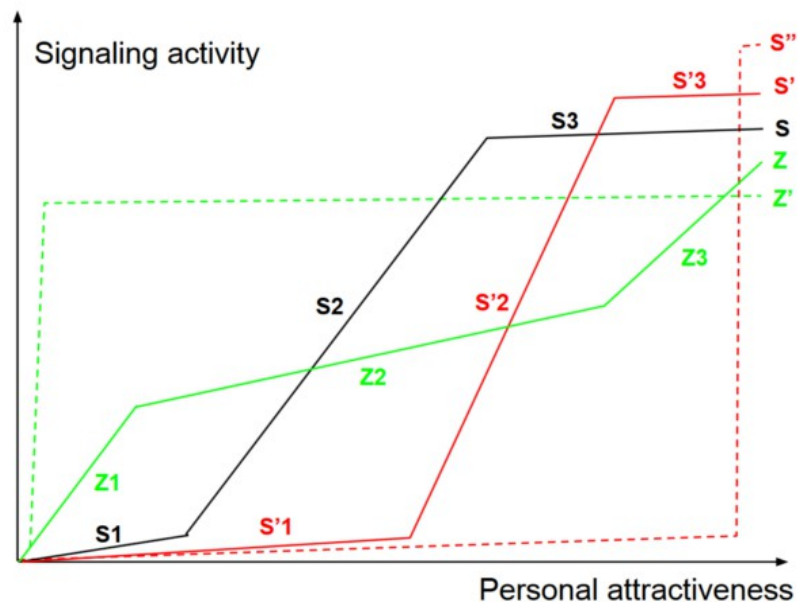

Four evolutionary scenarios of our current ESS (curve S black)
on the brink of the fourth singularity.
S' (red): Concentration of power induced by centralized networks.
S'' (dashed red): Centralization around some Artificial General Intelligence (AGI).
Z (green): Anoptical perspectives' acculturation towards a global immune system.
Z'' (dashed green): Mindplexes, Global Brain.

Scenario #1: The concentration of the economy in the hands of a few tends to transform the initial black curve S into the red one S'. That would mean:

> Section S'1: A bigger proportion of unattractive individuals with a poor social network and a low level of signalling activity.
> Section S'2: Much more competition between the individuals of this competitive class, which has proportionately a more limited size.
> Section S'3: Proportionally fewer individuals in this non-competitive class whose members are more attractive because they are fewer.

This transformation S->S' does not change the S shape of the curve. So there is no actual transformation of our ESS in that scenario, despite all the alleged 'disruptions' promised by the technology. Even if we extrapolate S' into S".

Scenario #2: S" suggests that some Artificial General Intelligence (AGI)—eventually controlled by some rare dominants considered as a semi-god—had taken over the whole human species and transformed all individuals into some asocial zombies. Say it is not really a desirable scenario, though some people dream about it as the only solution to solve human problems.

Scenario #3: This scenario requires making a bold assumption. Imagine that one day some scandals expose the disastrous ethical consequences of centralized networks and their illegitimate temporal perspective (the exploitation by a few of the emotions of the masses). Imagine that thereafter, some literacy about the anoptical perspectives begins to spread around the world driven by the vital need of the people to recover some autonomy. Imagine that these events trigger a tipping point where the second type of anoptical perspective (the digital perspective) attains the dominant position. Hence, S could mutate into Z, which has an inverse shape. That would mean:

> Section Z1: A quick socialization of individuals at the bottom of the ladder.
> Section Z2: A larger middle class with less competition among its members. Enjoy!
> Section Z3: A competitive elite would have replaced the non-competitive dominant individuals.

The inversion of shape S->Z would suggest that a slight change of our ESS has happened in order to restore the consistency and the autonomy of our society. One more time, our language would have prevailed upon the brute force and the misuse of our weaponry, especially the power of money.

Scenario #4: One can also extrapolate Z to Z'. Z' suggests the formation of some sort of Mindplexes where artificial intelligence and the human mind have combined with each other. Mindplexes would pave the way to the Global Brain, where radically new languages, for example like the one developed by Pierre Lévy (Lévy 2009)), would be practiced, between humans, and between humans and machines. According to Francis Heylighen, the Global Brain would pave the way to a new form of Eden (Heylighen 2014b).

This Z->Z' hypothesis would definitively put the world under the reign of the anoptical perspective of distributed networks. Its vanishing code would be some ACM or more generally a new social protocol. This may be seen as a new weapon that would enforce a new social order. In this case, it is not difficult to imagine that legitimacy issues would be very difficult because of the indescribable complexity of distributed networks. Only the acute and permanent awareness of all agents may prevent the possible takeover by some illegitimate systems or predatory artificial species. In other words, if one day, the Global Brain is a reality, it will need a Global Immune System to ensure its viability.

**Towards aethogenesis**

The evolutionary hypothesis I have explored disputes the admitted framework of the Darwinian theory, in which the notion of collective benefit is not included. Indeed, for Darwin and his successors, selective pressure rests upon the individuals only. However, Darwin himself noticed that the selective pressure on individuals can paradoxically lead to the selection of anti-selective behaviors. This 'inverse effect' (Tort, 1983) induces in fact a kind of selection of the social organizations considered as quasi-living species.

According to my hypothesis, human beings have specific cognitive resources that enable them to consider the perspectives created by their own technology. After having taken advantage of the optical perspectives for the development of our species, this would be the time of the anoptical ones. As network agents structured by the anoptical perspectives, we are (on the curve S) most often in the position of the 'selected', allowing them to shape our imagination and our judgments. But by becoming aware of these perspectives and by questioning their legitimacy, we could choose to evolve (towards the curves Z and Z') in order to be in the position of the 'selectors' of the quasi-living species to which we belong.

The Global Immune System, to which in this way we could participate, could select artificial species according to their legitimacy seen as an evolutionary advantage. Finally, it would be the relevant and proportionate update of our Evolutionarily Strategy. That could lead us to a new era, which I propose to call aethogenesis: The passage of a world without ethics to a world with ethics. The aethogenesis, as a fourth singularity, would trigger a new explosion of codes respecting the beings of all species, by beginning with our own one.